\documentclass[%
reprint,
superscriptaddress,
showpacs,
amsmath,amssymb,
aps,
prl,
longbibliography,
]{revtex4-1}

\usepackage{psfrag,graphicx,epsfig,color}
\usepackage{dcolumn}
\usepackage{bm}
\usepackage{natbib}
\usepackage[usenames,dvipsnames,svgnames,table]{xcolor}
\usepackage{subfigure}
\usepackage{rotating}
\usepackage{float}
\usepackage[nice]{nicefrac}

\def\re    {{R_\lambda}}

\begin{document}

\title{
Small-scale isotropy and ramp-cliff structures in scalar turbulence
}

\author{Dhawal Buaria }
\email[]{dhawal.buaria@nyu.edu}
\affiliation{Tandon School of Engineering, New York University, New York, NY 11201, USA}

\author{Matthew P. Clay}
\affiliation{School of Aerospace Engineering, Georgia Institute of Technology,
Atlanta, GA 30332, USA}
\author{Katepalli R. Sreenivasan}
\affiliation{Tandon School of Engineering, New York University, New York, NY 11201, USA}
\affiliation{Department of Physics and the Courant Institute of Mathematical Sciences,
New York University, New York, NY 10012, USA}

\author{P. K. Yeung}
\affiliation{School of Aerospace Engineering, Georgia Institute of Technology,
Atlanta, GA 30332, USA}
\affiliation{School of Mechanical Engineering, Georgia Institute of Technology,
Atlanta, GA 30332, USA}

\date{\today}


\begin{abstract}

Passive scalars advected by three-dimensional Navier-Stokes turbulence exhibit a fundamental anomaly in odd-order moments because of the characteristic ramp-cliff structures, violating small-scale isotropy. We use data from direct numerical simulations with grid resolution of up to $8192^3$ at high P\'eclet numbers to understand this anomaly as the scalar diffusivity, $D$, diminishes, or as the Schmidt number, $Sc = \nu/D$, increases; here $\nu$ is the kinematic viscosity of the fluid. The microscale Reynolds number varies from 140 to 650 and $Sc$ varies from 1 to 512. A simple model for the ramp-cliff structures is shown to characterize the scalar derivative statistics 
extremely well. It accurately captures how the small-scale isotropy is restored in the large-$Sc$ limit, and additionally suggests a slight correction to the Batchelor length scale as the relevant smallest scale in the scalar field.


\end{abstract}

\maketitle


\paragraph{Introduction:} The transport and mixing of a passive scalar by three-dimensional Navier-Stokes (NS) turbulence is an important problem in numerous natural and engineering processes \cite{Hill1976,ZW00,PD05}, and also fundamentally important because it is a candidate for applying the same ideas of universality as stem from Kolmogorov's seminal work on velocity fluctuations \cite{Corrsin51,MY.II,sreeni19}. An essential ingredient of this universality is that the anisotropies introduced by the forcing at large scales are ultimately lost at small scales, and increasingly smaller scales become increasingly isotropic \cite{MY.II}. A few decades of work has gone into showing that Kolmogorov's description is approximately valid for low-order statistics but breaks down for high-order quantities due to intermittency \cite{Frisch95,SA97,Ishihara09}. This breakdown stands out particularly for the scalar field, and manifests as a zeroth-order anomaly for odd-order moments of the derivative field: small-scale isotropy for the scalar requires odd-order derivative moments to vanish identically, whereas data from experiments and simulations show that the skewness (normalized third-order moment) in the direction of an imposed large-scale mean gradient remains to be of the order unity even at very high Reynolds numbers \cite{Mestayer76,krs77,KRS91,HS94,pumir94}, and that its sign correlates perfectly with the imposed mean gradient \cite{ST80}. 

This anomalous behavior, traced to the presence of ramp-cliff structures in the scalar field \cite{sab_79,KRS91,SS2000}, has been studied so far mostly when the Schmidt number $Sc=\nu/D = \mathcal{O}(1)$, where $\nu$ is the kinematic viscosity and $D$ is the diffusivity of the scalar. Earlier studies have indicated that the derivative skewness decreases as $Sc$ increases \cite{PK02,JS03,brethouwer03}, but the data, obtained at very low Reynolds numbers, were incidental to those papers. The question we answer in this Letter, utilizing data from state-of-the-art direct numerical simulations (DNS), is the nature of this change as $Sc$ increases; we also develop a physical model that provides excellent characterization of the data.

\begin{figure}
\begin{center}
\includegraphics[width=7.2cm]{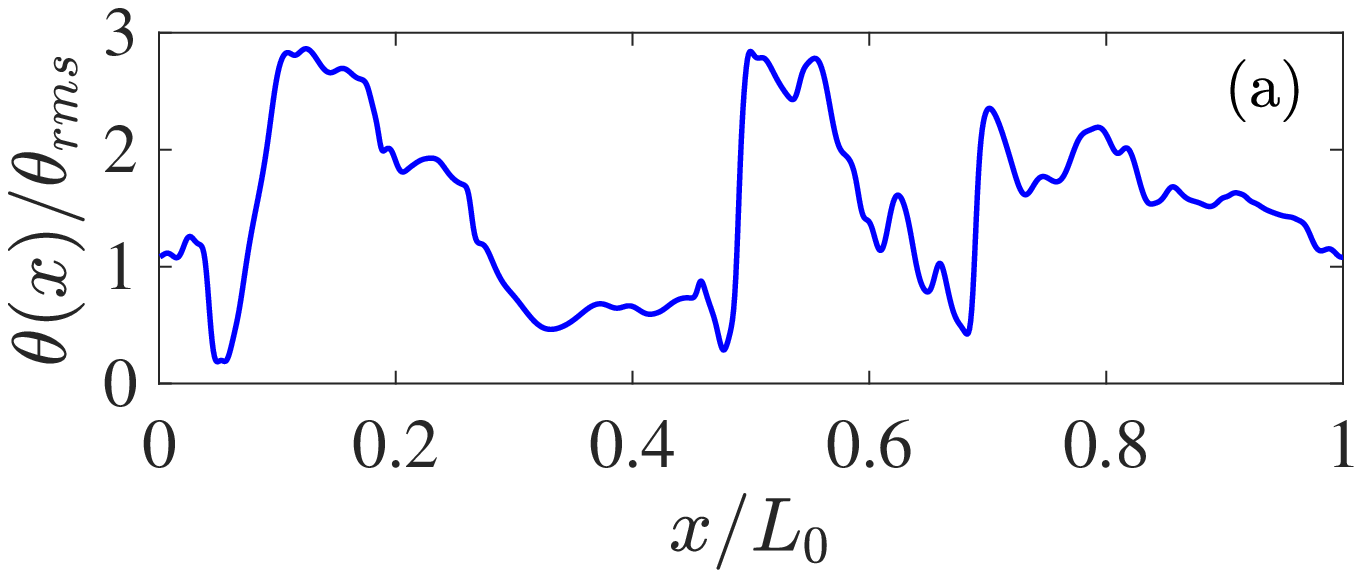} \\
\ \ \ \
\includegraphics[width=6.0cm]{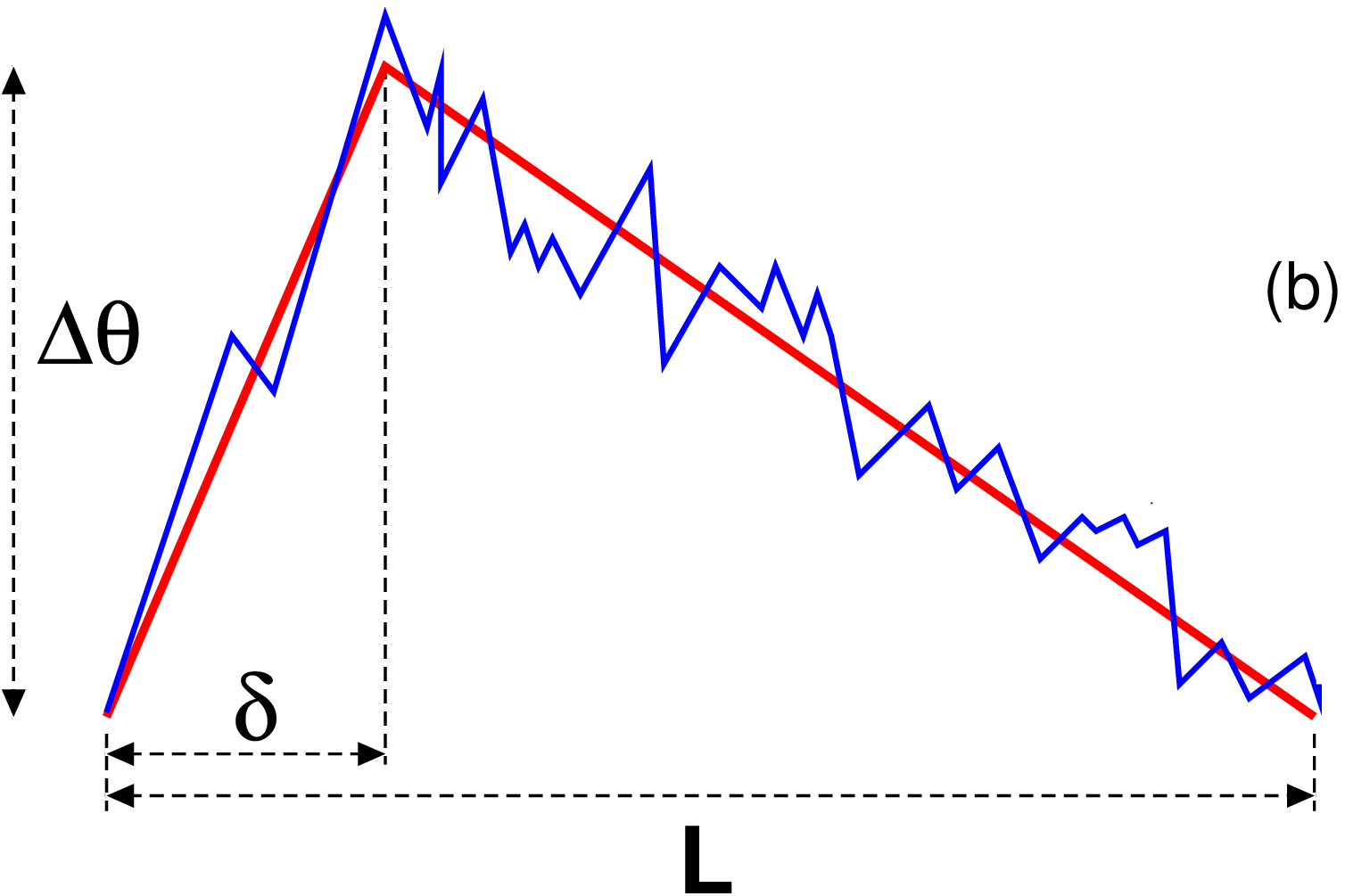}
\caption{
(a) Typical one-dimensional trace of the scalar field in the direction of the imposed mean-gradient ($x$), for $\re=140$ and $Sc = 1$, normalized by the rms value. $L_0=2\pi$ is the domain length. (b) A cartoon of the ramp-cliff model, based on the trace (but not to scale).
}
\label{fig:rc}
\end{center}
\end{figure}

\paragraph{DNS data:} The data examined in this work were generated using the canonical setup of isotropic turbulence in a periodic domain \cite{Ishihara09,BPB2020}, forced at large scales to maintain statistical stationarity. The passive scalar is obtained by simultaneously solving the advection-diffusion equation in the presence of mean uniform gradient $\nabla \Theta = (G,0,0)$ along one of the Cartesian directions, $x$ \cite{PK02}. The database utilized here is the same as in our recent work \cite{BCSY20}, and corresponds to microscale Reynolds number $\re\equiv u^\prime \lambda/\nu$ in the range $140-650$, where $u^\prime$ is the root-mean-square (rms) velocity fluctuation and $\lambda$ the Taylor microscale; $Sc$ lies in the range $1-512$. The P\'eclet number is the product $\re Sc$. As noted in \cite{BCSY20}, the data were generated using conventional Fourier pseudo-spectral methods for $Sc=1$, and a new hybrid approach for higher $Sc$ \cite{clay.cpc1, clay.cpc2, clay.omp}; in the latter method the velocity field was solved pseudo-spectrally while resolving the Kolmogorov length scale $\eta$, whereas the scalar field was solved using compact finite differences on a finer grid, so as to resolve the Batchelor scale $\eta_B =  \eta Sc^{-1/2}$. Because of the steep resolution requirements for $\eta_B$, earlier studies for high $Sc$ have been severely limited to low $\re$. Our database was generated using the largest grid sizes (of up to $8192^3$) currently feasible in DNS, and allows us to attain significantly higher $\re$ than before
\cite{BS2020,BBP2020}.

\begin{figure}
\begin{center}
\includegraphics[width=7.5cm]{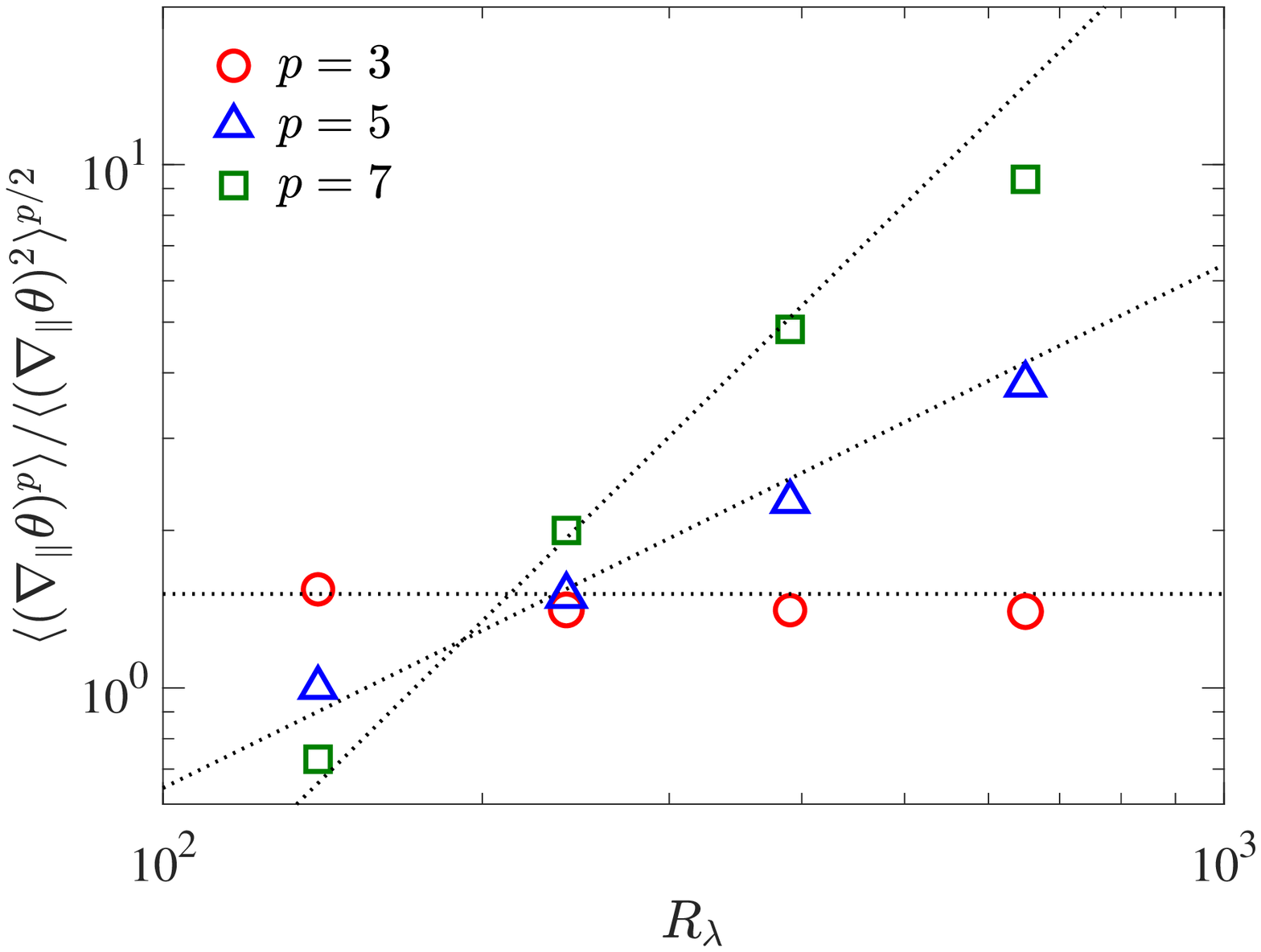} \\
\vspace{0.2cm}
\includegraphics[width=7.5cm]{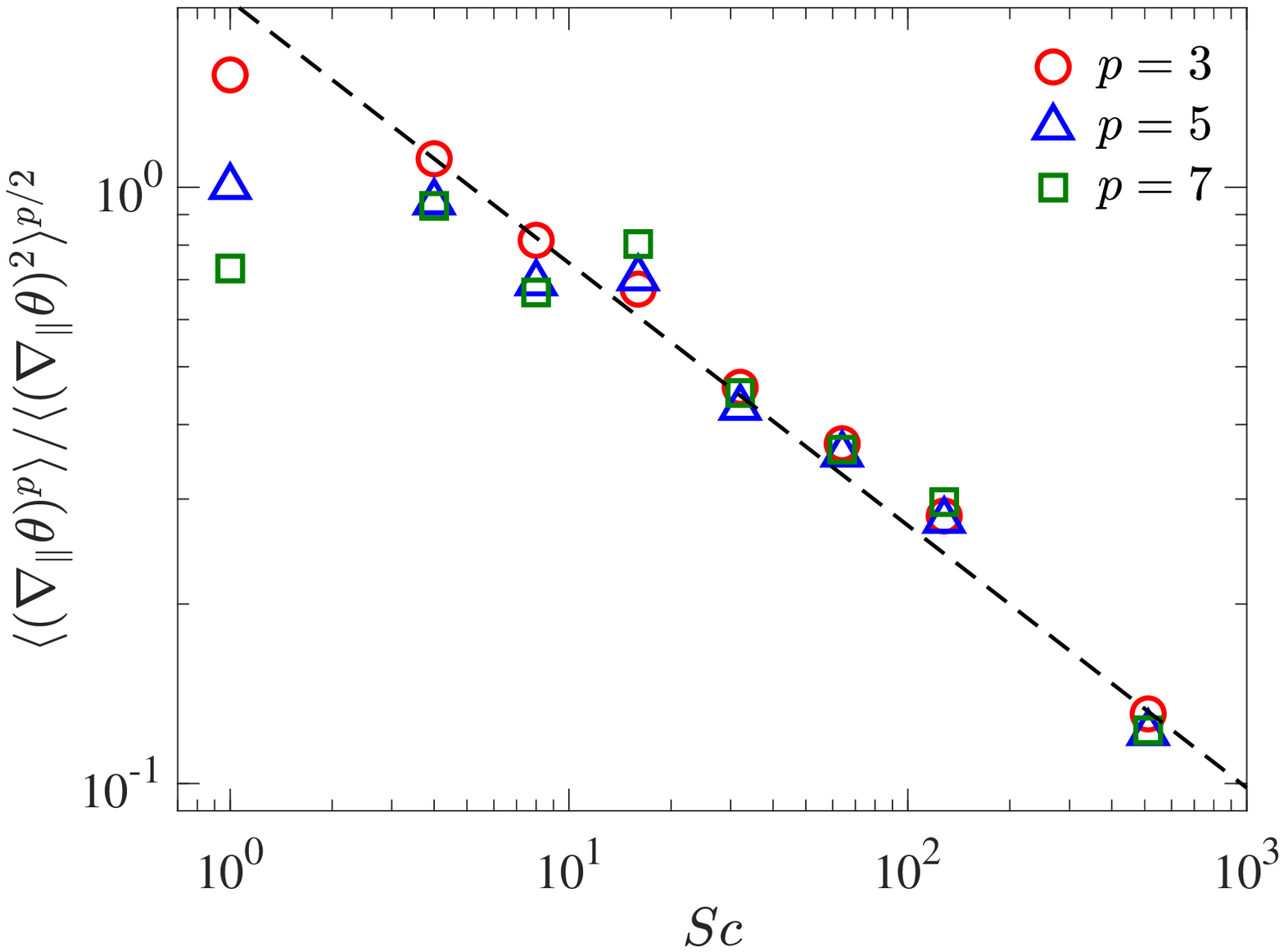}
\caption{
Normalized odd-order moments of the scalar derivative in the direction of the mean scalar gradient, (a) as a function of $\re$ at $Sc=1$, and (b) as a function of $Sc$ at $\re=140$. For clarity, the data for $p=5$ and $7$ are shifted down by factors of 100 and 16,000, respectively. The dotted lines in (a) correspond to power law slopes of $0,~1$ and $2$ (see Eq.~\eqref{eq:odd1}). The dashed line in (b) corresponds to a slope of $-0.45$ (instead of $-0.5$ given by Eq.~\eqref{eq:odd1}). Statistical errors are less than the symbol height but those resulting from finite grid resolution introduce some uncertainty in the seventh moment at $\re=650$.
}
\label{fig:cent}
\end{center}
\end{figure}

\paragraph{The ramp-cliff model:}
Figure \ref{fig:rc}a shows a typical trace of the scalar field for $Sc=1$, in which the characteristic ramp-cliff structures are clearly visible. 
Expectedly, the larger scalar gradients organized as sharp fronts (cliffs) are followed by regions of weaker gradients (ramps). 
A plausible physical reason for the ramp-cliff structure \cite{sreeni19} is the presence of coherent parcels of fluid with large scalar concentration values, moving at a finite velocity relative to the ambient, thus creating sharp fronts; the resulting ramp-cliff model for the scalar is shown in Fig.~\ref{fig:rc}b.
Its overall extent is on the order of the large scale $L$, whereas the cliff occurs over some small scale $\delta \ll L$, with the corresponding scalar increment $\Delta \theta$, which is of the order of the largest scalar fluctuation in the flow. Since the fluctuations of a scalar
advected in isotropic turbulence are Gaussian \cite{WG04}, it is reasonable to assume that $\Delta \theta \sim \theta_{rms}$, where $\theta_{rms}$ is the root-mean-square (rms) value \cite{pumir94b}. Thus, an odd moment of the scalar gradient would get its largest contribution from the cliff (with all other contributions essentially canceling each other). Given the gradient at the cliff is $\theta_{rms}/\delta$ and the fraction occupied by the cliff is $\delta/L$, one can write
\begin{align}
\left \langle \left( \nabla_\| \theta \right)^p  \right\rangle \sim 
\left(\frac{\theta_{rms}}{\delta}\right)^p  \times \frac{\delta}{L} \ ,
\label{eq:delta}
\end{align}
where $p >1$ is odd and $\nabla_\|\theta$ is the scalar derivative parallel to the imposed mean-gradient. Contributions of cliffs to even-order statistics can be regarded as small. 
It was argued in \cite{sreeni19} that $\delta \sim \eta_B$, the
Batchelor length scale, using which the following expression
for the standardized moments can
be derived:
\begin{align}
\frac{ \left \langle \left( \nabla_\| \theta \right)^p  \right\rangle }
{ \left \langle \left( \nabla_\| \theta \right)^2  \right\rangle^{p/2}} 
 \sim  Sc^{-1/2} \re^{(p-3)/2}  \ .
\label{eq:odd1}
\end{align}
The result for $p=3$ and $5$ was outlined in \cite{sreeni19}, but we have 
now generalized it to any arbitrary odd-order (see Supplementary for details).
The above derivation also assumes that the second derivative moment---which gives the mean scalar dissipation rate, $\langle \chi \rangle = 2D \langle |\nabla \theta|^2 \rangle$, can be written in terms of large-scale quantities, i.e., $\langle \chi \rangle \sim \theta_{rms}^2 u^\prime/L$, as naively anticipated from scalar
dissipation anomaly \cite{Donzis05,BYS.2016}. (However, we will show later
that this needs modification, since the scalar dissipation 
rate also has a weak $Sc$-dependence \cite{Donzis05,BCSY20}). The predictions of Eq.~\eqref{eq:odd1} are compared in Fig.~\ref{fig:cent} with the DNS data. Figure \ref{fig:cent}a shows that the normalized odd moments agree with expected variations on $\re$. Figure \ref{fig:cent}b shows that the $Sc$-variations are close to the prediction of the ramp model, but the best fit gives a slope of -0.45 instead of -0.5. 

\begin{figure}
\begin{center}
\includegraphics[width=7.2cm]{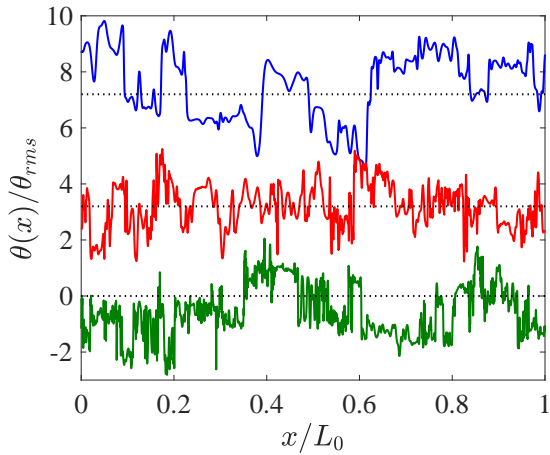}
\caption{
One-dimensional traces of the scalar field in the direction of the imposed mean gradient ($x$). Similar to Fig.~\ref{fig:rc}a, but for $Sc = 8,~64$ and $512$ (from top to bottom).
}
\label{fig:rc2}
\end{center}
\end{figure}

It is worth asking why the odd-order moments of the scalar derivative diminish with decreasing scalar diffusivity (i.e., increasing $Sc$). The reason can be seen briefly in the scalar traces for different values of $Sc$ (Fig.~\ref{fig:rc2}). The signals become more oscillatory as $Sc$ increases, and the underlying ramp structure, though present, makes smaller contributions to the overall derivative statistics. A related study can be found in \cite{BCSY20}.

\begin{figure}
\begin{center}
\includegraphics[width=7.5cm]{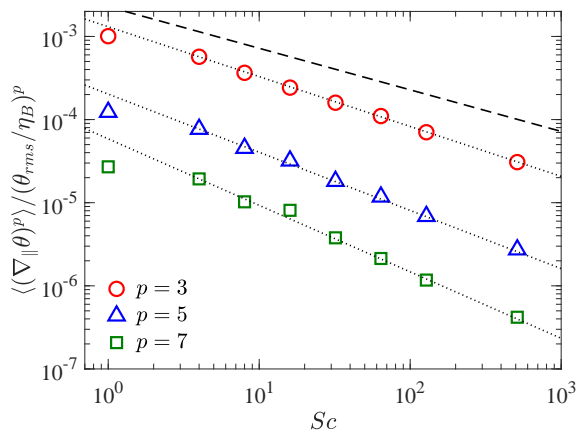} 
\caption{
The 3rd, 5th and 7th order moments of the scalar derivative in the direction of mean-gradient at $\re=140$, suitably normalized by $\theta_{rms}/\eta_B$. Dashed line corresponds to the $-1/2$-power predicted by Eq.~\eqref{eq:odd_etab}, whereas the dotted lines correspond to power laws with $-1/2 - \alpha(p-1)$, $\alpha = 0.05$ (see text). For clarity, the data for $p=5$ and $7$ are shifted down by factors of 4 and 16, respectively.
}
\label{fig:odd_sc}
\end{center}
\end{figure}

\paragraph{Refinement of the ramp-cliff model:} 
It is somewhat surprising that Eq.~\eqref{eq:odd1} of this elementary model agrees reasonably well with the data (Fig.~\ref{fig:cent}),  but it should be noted that the odd-order moments in Fig.~\ref{fig:cent} are normalized by the second moment. Recent studies have demonstrated that the second moment has a mild $Sc$-dependence \cite{BYS.2016,BCSY20}, so some cancellation of possible $Sc$-dependence between the second and odd-order moments aids the observed agreement. If, instead, we normalize the odd-order moments by the suitable power of the presumed gradient within the cliff, viz., $\theta_{rms}/\eta_B$, we get
\begin{align}
\frac{ \langle \left( \nabla_\| \theta \right)^p  \rangle}   
{\left( {\theta_{rms}} / {\eta_B} \right)^p} 
\sim Sc^{-1/2} \re^{-3/2}  \ .
\label{eq:odd_etab}
\end{align}
The model yields the same $Sc^{-1/2}$ dependence as Eq.~\eqref{eq:odd1} but simulations (Fig.~\ref{fig:odd_sc}) show increasing deviations from the $-1/2$ scaling as $p$ increases from 3 to 7.

Several possibilities to address these deviations can be considered, but the simplest is to take $\delta$ in Fig.~\ref{fig:rc}b as 
\begin{align}
\eta_D = \eta_B Sc^{\alpha} = \eta Sc^{-1/2 + \alpha} \ ,
\label{eq:alpha}
\end{align}
instead of $\eta_B$ as in \cite{sreeni19}; here $\alpha$ is a small positive number. 
Now using $\delta=\eta_D$ and substituting it in Eq.~\eqref{eq:delta} we find 
\begin{align}
\label{eq:beta}
\frac{ \left \langle \left( \nabla_\| \theta \right)^p  \right\rangle }
{\left( {\theta_{rms}} / {\eta_B} \right)^p }  
\sim & \ Sc^{\beta_p} \re^{-3/2}\ , \\  \nonumber
\ \ \ \ \text{with} & \ \ \ \beta_p = -\nicefrac{1}{2} - \alpha (p-1) \ ,
\end{align}
which provides an order-dependent scaling exponent. We now demonstrate the dynamical plausibility of this choice of $\alpha$ and address several concomitant issues.

\begin{figure}
\begin{center}
\includegraphics[width=7.5cm]{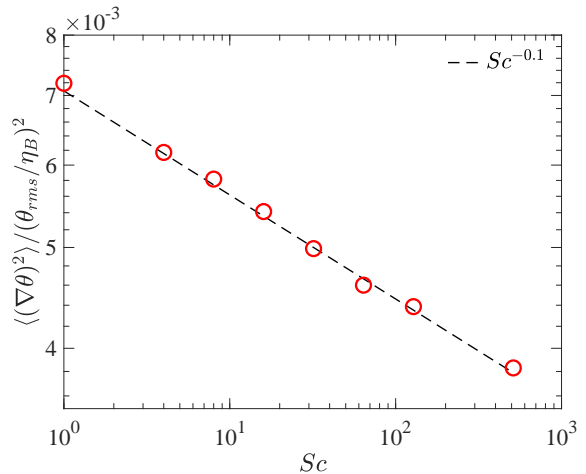}
\caption{
The variance of scalar gradient (which is also given as $\langle\chi\rangle/6D$) normalized by $\theta_{rms}/\eta_B$, $R_\lambda = 140$. The data can be well represented by a power law of the form $Sc^{-0.1}$.
}
\label{fig:scdiss}
\end{center}
\end{figure}

\paragraph{Justification for $\eta_D$ and dynamical consequences:}
An argument in favor of $\eta_D$ can be made in terms of the intermittency of energy dissipation \cite{MS91,BPBY2019}, which appears via $\eta$ in the definition $\eta_B  = \eta Sc^{-1/2}$, ultimately influencing the scalar field. It is thus reasonable to assume that $\delta$ in Fig.~\ref{fig:rc}b fluctuates around $\eta_B$ with an average value given by an $Sc$-dependent quantity such as $\eta_D$. 

Given that $\eta_D$ represents the dynamically smallest scale, it is natural to try to understand its influence on the even order moments. A known result from \cite{BSXDY,Donzis05} is that the normalized mean scalar dissipation rate decreases logarithmically with $Sc$. This result has been verified at high $\re$ in \cite{BYS.2016,BCSY20}, and can be written as
\begin{align}
\frac{\langle \chi \rangle L}{\theta_{rms}^2 u^\prime} \sim  
 c \ \frac{\re}{\log Sc} 
\end{align}
where $c$ is some constant, independent of $\re$. However, the $\log Sc$ dependence is only semi-empirical \cite{Donzis05}, and operationally indistinguishable from a weak power law dependence (which is more tractable for practical purposes). Since $\langle \chi \rangle$ is essentially the second moment of scalar derivatives, the above relation can be rewritten as
\begin{align}
\frac{\langle |\nabla \theta|^2 \rangle}{(\theta_{rms}/\eta_B)^2} \sim  
 c \  Sc^{-\gamma}  \ .
\label{eq:mom2}
\end{align}
where the $1/\log Sc$ dependence has been replaced $Sc^{-\gamma}$ (with $\gamma > 0$) and we have utilized classical scaling relations $L/\eta \sim \re^{3/2}$ and $\re^2 \sim Re = u^\prime L /\nu$ \cite{Frisch95}. We plot the left hand side of Eq.~\eqref{eq:mom2} versus $Sc$ in Fig.~\ref{fig:scdiss}, with the best fit giving $\gamma=0.1$. It now follows from the definition of $ \eta_D$ that $\alpha=\gamma/2=0.05$, allowing us to capture the Schmidt number scaling of the second moment. 
In fact, a similar consideration was also exploited in 
a recent work \cite{yasuda20}, where the authors also use the second moment of the scalar
derivative to define a Taylor length scale, which was then utilized to 
collapse many scalar statistics.

\begin{figure}
\begin{center}
\includegraphics[width=7.5cm]{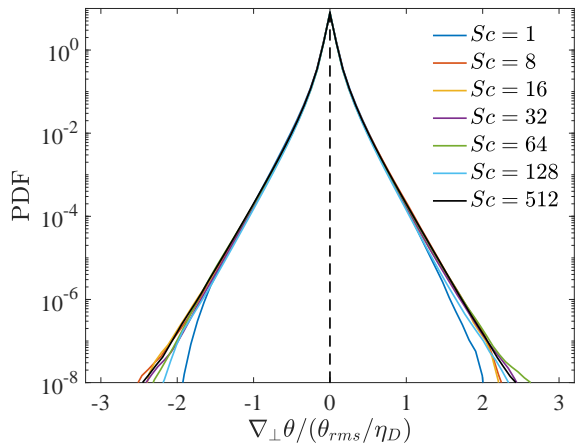}
\caption{
The PDFs of scalar derivative in the direction perpendicular to imposed mean-gradient, normalized by $\theta_{rms}/\eta_D$. $R_\lambda = 140$. The statistical error bars are negligible and the probability density is shown here only when the number of samples in the histogram was greater than $10^3$. The exponential tails of these dustributions was explored in \cite{ss94}.
}
\label{fig:pdfs_perp}
\end{center}
\end{figure}

The first outcome of this refinement is that it agrees very well with the data on odd-order moments (see dotted lines drawn in Fig.~\ref{fig:odd_sc}). In fact, combining the results from Eqs.~\eqref{eq:beta} and \eqref{eq:mom2} gives an $Sc^{-1/2+\alpha}$ relation for normalized odd-moments in Eq.~\eqref{eq:odd1}, which corrects the discrepancy noted in Fig.~\ref{fig:cent}b (the best slope being $-0.45$ instead of $-0.5$). As a second outcome, Fig.~\ref{fig:pdfs_perp} shows that the probability density functions (PDFs) of the scalar derivative perpendicular to the direction of the imposed mean-gradient show very good collapse for $Sc>1$ (with minor variation in extreme tails). 

\begin{figure}
\begin{center}
\includegraphics[width=7.5cm]{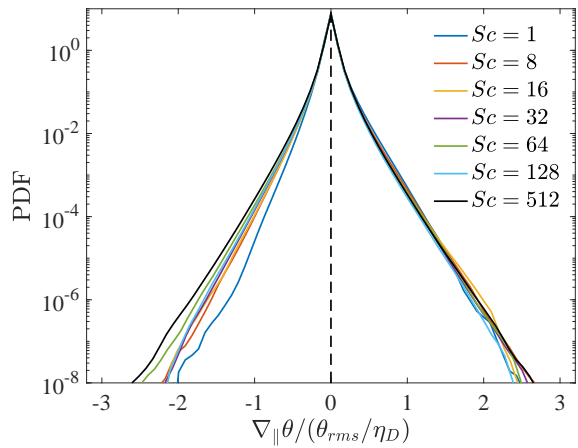}
\caption{
PDFs of scalar derivative parallel to the imposed mean gradient, normalized by $\theta_{rms}/\eta_D$. $R_\lambda = 140$.
}
\label{fig:pdfs}
\end{center}
\end{figure}

\begin{figure}
\begin{center}
\includegraphics[width=7.5cm]{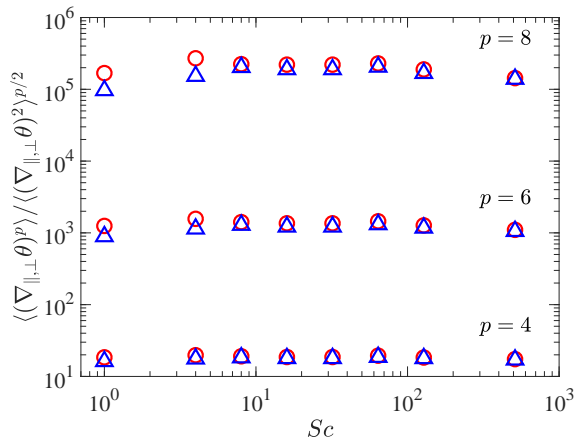}
\caption{
Moments of scalar derivative in parallel (red circles) and perpendicular (blue triangles) directions of the imposed mean gradient. With increasing $Sc$, the moments approach each other and are also independent of $Sc$, affirming the collapses of the PDFs seen in Figs.~\ref{fig:pdfs_perp} and \ref{fig:pdfs}.
}
\label{fig:even}
\end{center}
\end{figure}

A further outcome of using $\eta_D$ is that the positive sides of the PDFs of the scalar derivative parallel to the mean-gradient, corresponding essentially to the cliffs, collapse for all $Sc$ (see Fig.~\ref{fig:pdfs}). As $Sc$ increases, the left side of the PDF moves outwards rendering it symmetric for large $Sc$. Local isotropy dictates that the even moments of the scalar gradients both parallel and perpendicular to the imposed large mean gradients be equal, and, in fact, the high $Sc$-asymptote of the PDFs in Fig.~\ref{fig:pdfs} match the collapsed PDFs of Fig.~\ref{fig:pdfs_perp}. Thus, in the limit of large $Sc$, odd-order derivative moments in all directions are zero and even moments equal---in conformity with small-scale isotropy. The high-order even moments from both directions are explicitly compared in Fig.~\ref{fig:even}. It is seen that they approach each other and become independent of $Sc$ for $Sc \gtrsim 8$. 
Together these results consolidate the idea that the dynamically relevant smallest scale in the scalar field is $\eta_D$ (instead of $\eta_B$).

\paragraph{Summary:}
We have considered the important problem of the non-vanishing odd moments of the scalar derivative in the direction of the mean-gradient. This result violates the isotropy of small scales. We have shown that this feature can be accounted for by a simple mechanistic model for the ramp-cliff structure. This model predicts the normalized moments quite well. A closer look at the moments reveals certain departures from the model. These departures can be addressed by introducing a new diffusive scale that is different from the Batchelor scale. This new scale not only improves agreement with the data on odd-order moments but also allows us to collapse, for large $Sc$, all the PDFs of scalar gradients in all directions. In that limit, even moments of the derivative are equal, to all orders, in the direction of the mean gradient and perpendicular to it. In conclusion, our results provide a satisfactory characterization of all gradient statistics in passive scalar turbulence. It would be instructive to 
see how our results here, especially on the modification of
Batchelor length scale, translate to active scalars, such 
as temperature and salinity in the ocean.

\paragraph{Acknowledgments:} We thank Kartik Iyer and J\"org Schumacher for useful discussions. This research used resources of the Oak Ridge Leadership Computing Facility (OLCF), which is a Department of Energy (DOE) Office of Science user facility supported under Contract DE-AC05-00OR22725. We acknowledge the use of advanced computing resources at the OLCF under 2017 and 2018 INCITE Awards. Parts of the data analyzed in this work were obtained through National Science Foundation (NSF) Grant ACI-1036170, using resources of the Blue Waters sustained petascale computing project, which was supported by the NSF (awards OCI- 725070 and ACI-1238993) and the State of Illinois. DB also gratefully acknowledges the Gauss Centre for Supercomputing e.V. (www.gauss-centre.eu) for providing computing time on the supercomputer JUWELS at J\"ulich Supercomputing Centre, where the $Sc=1$ simulations were performed.



\begin{thebibliography}{36}%
\makeatletter
\providecommand \@ifxundefined [1]{%
 \@ifx{#1\undefined}
}%
\providecommand \@ifnum [1]{%
 \ifnum #1\expandafter \@firstoftwo
 \else \expandafter \@secondoftwo
 \fi
}%
\providecommand \@ifx [1]{%
 \ifx #1\expandafter \@firstoftwo
 \else \expandafter \@secondoftwo
 \fi
}%
\providecommand \natexlab [1]{#1}%
\providecommand \enquote  [1]{``#1''}%
\providecommand \bibnamefont  [1]{#1}%
\providecommand \bibfnamefont [1]{#1}%
\providecommand \citenamefont [1]{#1}%
\providecommand \href@noop [0]{\@secondoftwo}%
\providecommand \href [0]{\begingroup \@sanitize@url \@href}%
\providecommand \@href[1]{\@@startlink{#1}\@@href}%
\providecommand \@@href[1]{\endgroup#1\@@endlink}%
\providecommand \@sanitize@url [0]{\catcode `\\12\catcode `\$12\catcode
  `\&12\catcode `\#12\catcode `\^12\catcode `\_12\catcode `\%12\relax}%
\providecommand \@@startlink[1]{}%
\providecommand \@@endlink[0]{}%
\providecommand \url  [0]{\begingroup\@sanitize@url \@url }%
\providecommand \@url [1]{\endgroup\@href {#1}{\urlprefix }}%
\providecommand \urlprefix  [0]{URL }%
\providecommand \Eprint [0]{\href }%
\providecommand \doibase [0]{http://dx.doi.org/}%
\providecommand \selectlanguage [0]{\@gobble}%
\providecommand \bibinfo  [0]{\@secondoftwo}%
\providecommand \bibfield  [0]{\@secondoftwo}%
\providecommand \translation [1]{[#1]}%
\providecommand \BibitemOpen [0]{}%
\providecommand \bibitemStop [0]{}%
\providecommand \bibitemNoStop [0]{.\EOS\space}%
\providecommand \EOS [0]{\spacefactor3000\relax}%
\providecommand \BibitemShut  [1]{\csname bibitem#1\endcsname}%
\let\auto@bib@innerbib\@empty
\bibitem [{\citenamefont {Hill}(1976)}]{Hill1976}%
  \BibitemOpen
  \bibfield  {author} {\bibinfo {author} {\bibfnamefont {J.~C.}\ \bibnamefont
  {Hill}},\ }\bibfield  {title} {\enquote {\bibinfo {title} {Homogeneous
  turbulent mixing with chemical reaction},}\ }\href@noop {} {\bibfield
  {journal} {\bibinfo  {journal} {Annu.~Rev.~Fluid Mech.}\ }\textbf {\bibinfo
  {volume} {8}},\ \bibinfo {pages} {135--161} (\bibinfo {year}
  {1976})}\BibitemShut {NoStop}%
\bibitem [{\citenamefont {Warhaft}(2000)}]{ZW00}%
  \BibitemOpen
  \bibfield  {author} {\bibinfo {author} {\bibfnamefont {Z.}~\bibnamefont
  {Warhaft}},\ }\bibfield  {title} {\enquote {\bibinfo {title} {Passive scalars
  in turbulent flows},}\ }\href@noop {} {\bibfield  {journal} {\bibinfo
  {journal} {Annu. Rev. Fluid Mech.}\ }\textbf {\bibinfo {volume} {32}},\
  \bibinfo {pages} {203--240} (\bibinfo {year} {2000})}\BibitemShut {NoStop}%
\bibitem [{\citenamefont {Dimotakis}(2005)}]{PD05}%
  \BibitemOpen
  \bibfield  {author} {\bibinfo {author} {\bibfnamefont {P.~E.}\ \bibnamefont
  {Dimotakis}},\ }\bibfield  {title} {\enquote {\bibinfo {title} {Turbulent
  mixing},}\ }\href@noop {} {\bibfield  {journal} {\bibinfo  {journal} {Annu.
  Rev. Fluid Mech.}\ }\textbf {\bibinfo {volume} {37}},\ \bibinfo {pages}
  {329--356} (\bibinfo {year} {2005})}\BibitemShut {NoStop}%
\bibitem [{\citenamefont {Corrsin}(1951)}]{Corrsin51}%
  \BibitemOpen
  \bibfield  {author} {\bibinfo {author} {\bibfnamefont {S.}~\bibnamefont
  {Corrsin}},\ }\bibfield  {title} {\enquote {\bibinfo {title} {On the spectrum
  of isotropic temperature fluctuations in an isotropic turbulence},}\
  }\href@noop {} {\bibfield  {journal} {\bibinfo  {journal} {J. Appl. Phys.}\
  }\textbf {\bibinfo {volume} {22}},\ \bibinfo {pages} {469--473} (\bibinfo
  {year} {1951})}\BibitemShut {NoStop}%
\bibitem [{\citenamefont {Monin}\ and\ \citenamefont {Yaglom}(1975)}]{MY.II}%
  \BibitemOpen
  \bibfield  {author} {\bibinfo {author} {\bibfnamefont {A.~S.}\ \bibnamefont
  {Monin}}\ and\ \bibinfo {author} {\bibfnamefont {A.~M.}\ \bibnamefont
  {Yaglom}},\ }\href@noop {} {\emph {\bibinfo {title} {Statistical Fluid
  Mechanics}}},\ Vol.~\bibinfo {volume} {2}\ (\bibinfo  {publisher} {MIT
  Press},\ \bibinfo {year} {1975})\BibitemShut {NoStop}%
\bibitem [{\citenamefont {Sreenivasan}(2019)}]{sreeni19}%
  \BibitemOpen
  \bibfield  {author} {\bibinfo {author} {\bibfnamefont {K.~R.}\ \bibnamefont
  {Sreenivasan}},\ }\bibfield  {title} {\enquote {\bibinfo {title} {Turbulent
  mixing: A perspective},}\ }\href@noop {} {\bibfield  {journal} {\bibinfo
  {journal} {Proc.~Natl.~Acad.~Sci.}\ }\textbf {\bibinfo {volume} {116}},\
  \bibinfo {pages} {18175--18183} (\bibinfo {year} {2019})}\BibitemShut
  {NoStop}%
\bibitem [{\citenamefont {Frisch}(1995)}]{Frisch95}%
  \BibitemOpen
  \bibfield  {author} {\bibinfo {author} {\bibfnamefont {U.}~\bibnamefont
  {Frisch}},\ }\href@noop {} {\emph {\bibinfo {title} {Turbulence: the legacy
  of {Kolmogorov}}}}\ (\bibinfo  {publisher} {Cambridge University Press},\
  \bibinfo {address} {Cambridge},\ \bibinfo {year} {1995})\BibitemShut
  {NoStop}%
\bibitem [{\citenamefont {Sreenivasan}\ and\ \citenamefont
  {Antonia}(1997)}]{SA97}%
  \BibitemOpen
  \bibfield  {author} {\bibinfo {author} {\bibfnamefont {K.~S.}\ \bibnamefont
  {Sreenivasan}}\ and\ \bibinfo {author} {\bibfnamefont {R.~A.}\ \bibnamefont
  {Antonia}},\ }\bibfield  {title} {\enquote {\bibinfo {title} {The
  phenomenology of small-scale turbulence},}\ }\href@noop {} {\bibfield
  {journal} {\bibinfo  {journal} {Annu.~Rev.~Fluid~Mech.}\ }\textbf {\bibinfo
  {volume} {29}},\ \bibinfo {pages} {435--77} (\bibinfo {year}
  {1997})}\BibitemShut {NoStop}%
\bibitem [{\citenamefont {Ishihara}\ \emph {et~al.}(2009)\citenamefont
  {Ishihara}, \citenamefont {Gotoh},\ and\ \citenamefont
  {Kaneda}}]{Ishihara09}%
  \BibitemOpen
  \bibfield  {author} {\bibinfo {author} {\bibfnamefont {T.}~\bibnamefont
  {Ishihara}}, \bibinfo {author} {\bibfnamefont {T.}~\bibnamefont {Gotoh}}, \
  and\ \bibinfo {author} {\bibfnamefont {Y.}~\bibnamefont {Kaneda}},\
  }\bibfield  {title} {\enquote {\bibinfo {title} {Study of high-{Reynolds}
  number isotropic turbulence by direct numerical simulations},}\ }\href@noop
  {} {\bibfield  {journal} {\bibinfo  {journal} {Ann.~Rev.~Fluid~Mech.}\
  }\textbf {\bibinfo {volume} {41}},\ \bibinfo {pages} {165--80} (\bibinfo
  {year} {2009})}\BibitemShut {NoStop}%
\bibitem [{\citenamefont {Mestayer}\ \emph {et~al.}(1976)\citenamefont
  {Mestayer}, \citenamefont {Gibson}, \citenamefont {Coantic},\ and\
  \citenamefont {Patel}}]{Mestayer76}%
  \BibitemOpen
  \bibfield  {author} {\bibinfo {author} {\bibfnamefont {P.~G.}\ \bibnamefont
  {Mestayer}}, \bibinfo {author} {\bibfnamefont {C.~H.}\ \bibnamefont
  {Gibson}}, \bibinfo {author} {\bibfnamefont {M.~F.}\ \bibnamefont {Coantic}},
  \ and\ \bibinfo {author} {\bibfnamefont {A.~S.}\ \bibnamefont {Patel}},\
  }\bibfield  {title} {\enquote {\bibinfo {title} {Local anisotropy in heated
  and cooled turbulent boundary layers},}\ }\href@noop {} {\bibfield  {journal}
  {\bibinfo  {journal} {Phys. Fluids}\ }\textbf {\bibinfo {volume} {19}},\
  \bibinfo {pages} {1279--1287} (\bibinfo {year} {1976})}\BibitemShut {NoStop}%
\bibitem [{\citenamefont {Sreenivasan}\ and\ \citenamefont
  {Antonia}(1977)}]{krs77}%
  \BibitemOpen
  \bibfield  {author} {\bibinfo {author} {\bibfnamefont {K.~R.}\ \bibnamefont
  {Sreenivasan}}\ and\ \bibinfo {author} {\bibfnamefont {R.~A.}\ \bibnamefont
  {Antonia}},\ }\bibfield  {title} {\enquote {\bibinfo {title} {Skewness of
  temperature derivatives in turbulent shear flows},}\ }\href@noop {}
  {\bibfield  {journal} {\bibinfo  {journal} {Phys. Fluids}\ }\textbf {\bibinfo
  {volume} {20}},\ \bibinfo {pages} {1986--1988} (\bibinfo {year}
  {1977})}\BibitemShut {NoStop}%
\bibitem [{\citenamefont {Sreenivasan}(1991)}]{KRS91}%
  \BibitemOpen
  \bibfield  {author} {\bibinfo {author} {\bibfnamefont {K.~R.}\ \bibnamefont
  {Sreenivasan}},\ }\bibfield  {title} {\enquote {\bibinfo {title} {On local
  isotropy of passive scalars in turbulent shear flows},}\ }\href@noop {}
  {\bibfield  {journal} {\bibinfo  {journal} {Proc. R. Soc. Lond. A}\ }\textbf
  {\bibinfo {volume} {434}},\ \bibinfo {pages} {165--182} (\bibinfo {year}
  {1991})}\BibitemShut {NoStop}%
\bibitem [{\citenamefont {Holzer}\ and\ \citenamefont {Siggia}(1994)}]{HS94}%
  \BibitemOpen
  \bibfield  {author} {\bibinfo {author} {\bibfnamefont {M.}~\bibnamefont
  {Holzer}}\ and\ \bibinfo {author} {\bibfnamefont {E.~D.}\ \bibnamefont
  {Siggia}},\ }\bibfield  {title} {\enquote {\bibinfo {title} {Turbulent mixing
  of a passive scalar},}\ }\href@noop {} {\bibfield  {journal} {\bibinfo
  {journal} {Phys. Fluids}\ }\textbf {\bibinfo {volume} {6}},\ \bibinfo {pages}
  {1820--1837} (\bibinfo {year} {1994})}\BibitemShut {NoStop}%
\bibitem [{\citenamefont {Pumir}(1994{\natexlab{a}})}]{pumir94}%
  \BibitemOpen
  \bibfield  {author} {\bibinfo {author} {\bibfnamefont {A.}~\bibnamefont
  {Pumir}},\ }\bibfield  {title} {\enquote {\bibinfo {title} {A numerical study
  of the mixing of a passive scalar in three dimensions in the presence of a
  mean gradient},}\ }\href@noop {} {\bibfield  {journal} {\bibinfo  {journal}
  {Phys. Fluids}\ }\textbf {\bibinfo {volume} {6}},\ \bibinfo {pages}
  {2118--2132} (\bibinfo {year} {1994}{\natexlab{a}})}\BibitemShut {NoStop}%
\bibitem [{\citenamefont {Sreenivasan}\ and\ \citenamefont
  {Tavoularis}(1980)}]{ST80}%
  \BibitemOpen
  \bibfield  {author} {\bibinfo {author} {\bibfnamefont {K.~R.}\ \bibnamefont
  {Sreenivasan}}\ and\ \bibinfo {author} {\bibfnamefont {S.}~\bibnamefont
  {Tavoularis}},\ }\bibfield  {title} {\enquote {\bibinfo {title} {On the
  skewness of the temperature derivative in turbulent flows},}\ }\href@noop {}
  {\bibfield  {journal} {\bibinfo  {journal} {J.~Fluid Mech.}\ }\textbf
  {\bibinfo {volume} {101}},\ \bibinfo {pages} {783--795} (\bibinfo {year}
  {1980})}\BibitemShut {NoStop}%
\bibitem [{\citenamefont {Sreenivasan}\ \emph {et~al.}(1979)\citenamefont
  {Sreenivasan}, \citenamefont {Antonia},\ and\ \citenamefont
  {Britz}}]{sab_79}%
  \BibitemOpen
  \bibfield  {author} {\bibinfo {author} {\bibfnamefont {K.~R.}\ \bibnamefont
  {Sreenivasan}}, \bibinfo {author} {\bibfnamefont {R.~A.}\ \bibnamefont
  {Antonia}}, \ and\ \bibinfo {author} {\bibfnamefont {D.}~\bibnamefont
  {Britz}},\ }\bibfield  {title} {\enquote {\bibinfo {title} {Local isotropy
  and large structures in a heated turbulent jet},}\ }\href@noop {} {\bibfield
  {journal} {\bibinfo  {journal} {J. Fluid Mech.}\ }\textbf {\bibinfo {volume}
  {94}},\ \bibinfo {pages} {745–775} (\bibinfo {year} {1979})}\BibitemShut
  {NoStop}%
\bibitem [{\citenamefont {Shraiman}\ and\ \citenamefont
  {Siggia}(2000)}]{SS2000}%
  \BibitemOpen
  \bibfield  {author} {\bibinfo {author} {\bibfnamefont {B.~I.}\ \bibnamefont
  {Shraiman}}\ and\ \bibinfo {author} {\bibfnamefont {E.~D.}\ \bibnamefont
  {Siggia}},\ }\bibfield  {title} {\enquote {\bibinfo {title} {Scalar
  turbulence},}\ }\href@noop {} {\bibfield  {journal} {\bibinfo  {journal}
  {Nature}\ }\textbf {\bibinfo {volume} {405}},\ \bibinfo {pages} {639--646}
  (\bibinfo {year} {2000})}\BibitemShut {NoStop}%
\bibitem [{\citenamefont {Yeung}\ \emph {et~al.}(2002)\citenamefont {Yeung},
  \citenamefont {Xu},\ and\ \citenamefont {Sreenivasan}}]{PK02}%
  \BibitemOpen
  \bibfield  {author} {\bibinfo {author} {\bibfnamefont {P.~K.}\ \bibnamefont
  {Yeung}}, \bibinfo {author} {\bibfnamefont {S.}~\bibnamefont {Xu}}, \ and\
  \bibinfo {author} {\bibfnamefont {K.~R.}\ \bibnamefont {Sreenivasan}},\
  }\bibfield  {title} {\enquote {\bibinfo {title} {{S}chmidt number effects on
  turbulent transport with uniform mean scalar gradient},}\ }\href@noop {}
  {\bibfield  {journal} {\bibinfo  {journal} {Phys. Fluids}\ }\textbf {\bibinfo
  {volume} {14}},\ \bibinfo {pages} {4178--4191} (\bibinfo {year}
  {2002})}\BibitemShut {NoStop}%
\bibitem [{\citenamefont {Schumacher}\ and\ \citenamefont
  {Sreenivasan}(2003)}]{JS03}%
  \BibitemOpen
  \bibfield  {author} {\bibinfo {author} {\bibfnamefont {J.}~\bibnamefont
  {Schumacher}}\ and\ \bibinfo {author} {\bibfnamefont {K.~R.}\ \bibnamefont
  {Sreenivasan}},\ }\bibfield  {title} {\enquote {\bibinfo {title} {Geometric
  features of the mixing of passive scalars at high {S}chmidt numbers},}\
  }\href@noop {} {\bibfield  {journal} {\bibinfo  {journal} {Phys. Rev. Lett.}\
  }\textbf {\bibinfo {volume} {91}},\ \bibinfo {pages} {174501} (\bibinfo
  {year} {2003})}\BibitemShut {NoStop}%
\bibitem [{\citenamefont {Brethouwer}\ \emph {et~al.}(2003)\citenamefont
  {Brethouwer}, \citenamefont {Hunt},\ and\ \citenamefont
  {Nieuwstadt}}]{brethouwer03}%
  \BibitemOpen
  \bibfield  {author} {\bibinfo {author} {\bibfnamefont {G.}~\bibnamefont
  {Brethouwer}}, \bibinfo {author} {\bibfnamefont {J.~C.~R.}\ \bibnamefont
  {Hunt}}, \ and\ \bibinfo {author} {\bibfnamefont {F.~T.~M.}\ \bibnamefont
  {Nieuwstadt}},\ }\bibfield  {title} {\enquote {\bibinfo {title}
  {Micro-structure and {Lagrangian} statistics of the scalar field with a mean
  gradient in isotropic turbulence},}\ }\href@noop {} {\bibfield  {journal}
  {\bibinfo  {journal} {J.~Fluid Mech.}\ }\textbf {\bibinfo {volume} {474}},\
  \bibinfo {pages} {193--225} (\bibinfo {year} {2003})}\BibitemShut {NoStop}%
\bibitem [{\citenamefont {Buaria}\ \emph
  {et~al.}(2020{\natexlab{a}})\citenamefont {Buaria}, \citenamefont {Pumir},\
  and\ \citenamefont {Bodenschatz}}]{BPB2020}%
  \BibitemOpen
  \bibfield  {author} {\bibinfo {author} {\bibfnamefont {D.}~\bibnamefont
  {Buaria}}, \bibinfo {author} {\bibfnamefont {A.}~\bibnamefont {Pumir}}, \
  and\ \bibinfo {author} {\bibfnamefont {E.}~\bibnamefont {Bodenschatz}},\
  }\bibfield  {title} {\enquote {\bibinfo {title} {Self-attenuation of extreme
  events in {Navier-Stokes} turbulence},}\ }\href@noop {} {\bibfield  {journal}
  {\bibinfo  {journal} {Nat. Commun.}\ }\textbf {\bibinfo {volume} {11}},\
  \bibinfo {pages} {5852} (\bibinfo {year} {2020}{\natexlab{a}})}\BibitemShut
  {NoStop}%
\bibitem [{\citenamefont {Buaria}\ \emph {et~al.}()\citenamefont {Buaria},
  \citenamefont {Clay}, \citenamefont {Sreenivasan},\ and\ \citenamefont
  {Yeung}}]{BCSY20}%
  \BibitemOpen
  \bibfield  {author} {\bibinfo {author} {\bibfnamefont {D.}~\bibnamefont
  {Buaria}}, \bibinfo {author} {\bibfnamefont {M.~P.}\ \bibnamefont {Clay}},
  \bibinfo {author} {\bibfnamefont {K.~R.}\ \bibnamefont {Sreenivasan}}, \ and\
  \bibinfo {author} {\bibfnamefont {P.~K.}\ \bibnamefont {Yeung}},\ }\bibfield
  {title} {\enquote {\bibinfo {title} {Turbulence is an ineffective mixer when
  schmidt numbers are large},}\ }\href@noop {} {\bibfield  {journal} {\bibinfo
  {journal} {Phys.~Rev.~Lett.}\ }\textbf {\bibinfo {volume} {126}},\ \bibinfo
  {pages} {074501}}\BibitemShut {NoStop}%
\bibitem [{\citenamefont {Clay}\ \emph
  {et~al.}(2017{\natexlab{a}})\citenamefont {Clay}, \citenamefont {Buaria},
  \citenamefont {Gotoh},\ and\ \citenamefont {Yeung}}]{clay.cpc1}%
  \BibitemOpen
  \bibfield  {author} {\bibinfo {author} {\bibfnamefont {M.~P.}\ \bibnamefont
  {Clay}}, \bibinfo {author} {\bibfnamefont {D.}~\bibnamefont {Buaria}},
  \bibinfo {author} {\bibfnamefont {T.}~\bibnamefont {Gotoh}}, \ and\ \bibinfo
  {author} {\bibfnamefont {P.~K.}\ \bibnamefont {Yeung}},\ }\bibfield  {title}
  {\enquote {\bibinfo {title} {A dual communicator and dual grid-resolution
  algorithm for petascale simulations of turbulent mixing at high {Schmidt}
  number},}\ }\href@noop {} {\bibfield  {journal} {\bibinfo  {journal} {Comput.
  Phys. Commun.}\ }\textbf {\bibinfo {volume} {219}},\ \bibinfo {pages}
  {313--328} (\bibinfo {year} {2017}{\natexlab{a}})}\BibitemShut {NoStop}%
\bibitem [{\citenamefont {Clay}\ \emph {et~al.}(2018)\citenamefont {Clay},
  \citenamefont {Buaria}, \citenamefont {Yeung},\ and\ \citenamefont
  {Gotoh}}]{clay.cpc2}%
  \BibitemOpen
  \bibfield  {author} {\bibinfo {author} {\bibfnamefont {M.~P.}\ \bibnamefont
  {Clay}}, \bibinfo {author} {\bibfnamefont {D.}~\bibnamefont {Buaria}},
  \bibinfo {author} {\bibfnamefont {P.~K.}\ \bibnamefont {Yeung}}, \ and\
  \bibinfo {author} {\bibfnamefont {T.}~\bibnamefont {Gotoh}},\ }\bibfield
  {title} {\enquote {\bibinfo {title} {{GPU} acceleration of a petascale
  application for turbulent mixing at high {Schmidt} number using {OpenMP}
  4.5},}\ }\href@noop {} {\bibfield  {journal} {\bibinfo  {journal} {Comput.
  Phys. Commun.}\ }\textbf {\bibinfo {volume} {228}},\ \bibinfo {pages}
  {100--114} (\bibinfo {year} {2018})}\BibitemShut {NoStop}%
\bibitem [{\citenamefont {Clay}\ \emph
  {et~al.}(2017{\natexlab{b}})\citenamefont {Clay}, \citenamefont {Buaria},\
  and\ \citenamefont {Yeung}}]{clay.omp}%
  \BibitemOpen
  \bibfield  {author} {\bibinfo {author} {\bibfnamefont {M.~P.}\ \bibnamefont
  {Clay}}, \bibinfo {author} {\bibfnamefont {D.}~\bibnamefont {Buaria}}, \ and\
  \bibinfo {author} {\bibfnamefont {P.~K.}\ \bibnamefont {Yeung}},\ }\bibfield
  {title} {\enquote {\bibinfo {title} {Improving scalability and accelerating
  petascale turbulence simulatio using {OpenMP}},}\ }in\ \href@noop {} {\emph
  {\bibinfo {booktitle} {Proceedings of {OpenMP} Conference}}}\ (\bibinfo
  {address} {Stony Brook University, NY},\ \bibinfo {year} {2017})\BibitemShut
  {NoStop}%
\bibitem [{\citenamefont {Buaria}\ and\ \citenamefont
  {Sreenivasan}(2020)}]{BS2020}%
  \BibitemOpen
  \bibfield  {author} {\bibinfo {author} {\bibfnamefont {D.}~\bibnamefont
  {Buaria}}\ and\ \bibinfo {author} {\bibfnamefont {K.~R.}\ \bibnamefont
  {Sreenivasan}},\ }\bibfield  {title} {\enquote {\bibinfo {title} {Dissipation
  range of the energy spectrum in high {Reynolds} number turbulence},}\
  }\href@noop {} {\bibfield  {journal} {\bibinfo  {journal}
  {Phys.~Rev.~Fluids}\ }\textbf {\bibinfo {volume} {5}},\ \bibinfo {pages}
  {092601(R)} (\bibinfo {year} {2020})}\BibitemShut {NoStop}%
\bibitem [{\citenamefont {Buaria}\ \emph
  {et~al.}(2020{\natexlab{b}})\citenamefont {Buaria}, \citenamefont
  {Bodenschatz},\ and\ \citenamefont {Pumir}}]{BBP2020}%
  \BibitemOpen
  \bibfield  {author} {\bibinfo {author} {\bibfnamefont {D.}~\bibnamefont
  {Buaria}}, \bibinfo {author} {\bibfnamefont {E.}~\bibnamefont {Bodenschatz}},
  \ and\ \bibinfo {author} {\bibfnamefont {A.}~\bibnamefont {Pumir}},\
  }\bibfield  {title} {\enquote {\bibinfo {title} {Vortex stretching and
  enstrophy production in high {Reynolds} number turbulence},}\ }\href@noop {}
  {\bibfield  {journal} {\bibinfo  {journal} {Phys.~Rev.~Fluids}\ }\textbf
  {\bibinfo {volume} {5}},\ \bibinfo {pages} {104602} (\bibinfo {year}
  {2020}{\natexlab{b}})}\BibitemShut {NoStop}%
\bibitem [{\citenamefont {Watanabe}\ and\ \citenamefont {Gotoh}(2004)}]{WG04}%
  \BibitemOpen
  \bibfield  {author} {\bibinfo {author} {\bibfnamefont {T.}~\bibnamefont
  {Watanabe}}\ and\ \bibinfo {author} {\bibfnamefont {T.}~\bibnamefont
  {Gotoh}},\ }\bibfield  {title} {\enquote {\bibinfo {title} {Statistics of a
  passive scalar in homogeneous turbulence},}\ }\href@noop {} {\bibfield
  {journal} {\bibinfo  {journal} {New J. Phys.}\ }\textbf {\bibinfo {volume}
  {6}},\ \bibinfo {pages} {40} (\bibinfo {year} {2004})}\BibitemShut {NoStop}%
\bibitem [{\citenamefont {Pumir}(1994{\natexlab{b}})}]{pumir94b}%
  \BibitemOpen
  \bibfield  {author} {\bibinfo {author} {\bibfnamefont {A.}~\bibnamefont
  {Pumir}},\ }\bibfield  {title} {\enquote {\bibinfo {title} {Small-scale
  properties of scalar and velocity differences in three-dimensional
  turbulence},}\ }\href@noop {} {\bibfield  {journal} {\bibinfo  {journal}
  {Phys.~Fluids}\ }\textbf {\bibinfo {volume} {6}},\ \bibinfo {pages}
  {3974--3984} (\bibinfo {year} {1994}{\natexlab{b}})}\BibitemShut {NoStop}%
\bibitem [{\citenamefont {Donzis}\ \emph {et~al.}(2005)\citenamefont {Donzis},
  \citenamefont {Sreenivasan},\ and\ \citenamefont {Yeung}}]{Donzis05}%
  \BibitemOpen
  \bibfield  {author} {\bibinfo {author} {\bibfnamefont {D.~A.}\ \bibnamefont
  {Donzis}}, \bibinfo {author} {\bibfnamefont {K.~R.}\ \bibnamefont
  {Sreenivasan}}, \ and\ \bibinfo {author} {\bibfnamefont {P.~K.}\ \bibnamefont
  {Yeung}},\ }\bibfield  {title} {\enquote {\bibinfo {title} {Scalar
  dissipation rate and dissipative anomaly in isotropic turbulence},}\
  }\href@noop {} {\bibfield  {journal} {\bibinfo  {journal} {J.~Fluid Mech.}\
  }\textbf {\bibinfo {volume} {532}},\ \bibinfo {pages} {199--216} (\bibinfo
  {year} {2005})}\BibitemShut {NoStop}%
\bibitem [{\citenamefont {Buaria}\ \emph {et~al.}(2016)\citenamefont {Buaria},
  \citenamefont {Yeung},\ and\ \citenamefont {Sawford}}]{BYS.2016}%
  \BibitemOpen
  \bibfield  {author} {\bibinfo {author} {\bibfnamefont {D.}~\bibnamefont
  {Buaria}}, \bibinfo {author} {\bibfnamefont {P.~K.}\ \bibnamefont {Yeung}}, \
  and\ \bibinfo {author} {\bibfnamefont {B.~L.}\ \bibnamefont {Sawford}},\
  }\bibfield  {title} {\enquote {\bibinfo {title} {{A Lagrangian} study of
  turbulent mixing: forward and backward dispersion of molecular trajectories
  in isotropic turbulence},}\ }\href@noop {} {\bibfield  {journal} {\bibinfo
  {journal} {J.~Fluid Mech.}\ }\textbf {\bibinfo {volume} {{799}}},\ \bibinfo
  {pages} {{352--382}} (\bibinfo {year} {2016})}\BibitemShut {NoStop}%
\bibitem [{\citenamefont {Meneveau}\ and\ \citenamefont
  {Sreenivasan}(1991)}]{MS91}%
  \BibitemOpen
  \bibfield  {author} {\bibinfo {author} {\bibfnamefont {C.}~\bibnamefont
  {Meneveau}}\ and\ \bibinfo {author} {\bibfnamefont {K.~R.}\ \bibnamefont
  {Sreenivasan}},\ }\bibfield  {title} {\enquote {\bibinfo {title} {The
  multifractal nature of turbulent energy dissipation},}\ }\href@noop {}
  {\bibfield  {journal} {\bibinfo  {journal} {J.~Fluid Mech.}\ }\textbf
  {\bibinfo {volume} {224}},\ \bibinfo {pages} {429–--484} (\bibinfo {year}
  {1991})}\BibitemShut {NoStop}%
\bibitem [{\citenamefont {Buaria}\ \emph {et~al.}(2019)\citenamefont {Buaria},
  \citenamefont {Pumir}, \citenamefont {Bodenschatz},\ and\ \citenamefont
  {Yeung}}]{BPBY2019}%
  \BibitemOpen
  \bibfield  {author} {\bibinfo {author} {\bibfnamefont {D.}~\bibnamefont
  {Buaria}}, \bibinfo {author} {\bibfnamefont {A.}~\bibnamefont {Pumir}},
  \bibinfo {author} {\bibfnamefont {E.}~\bibnamefont {Bodenschatz}}, \ and\
  \bibinfo {author} {\bibfnamefont {P.~K.}\ \bibnamefont {Yeung}},\ }\bibfield
  {title} {\enquote {\bibinfo {title} {Extreme velocity gradients in turbulent
  flows},}\ }\href@noop {} {\bibfield  {journal} {\bibinfo  {journal} {New
  J.~Phys.}\ }\textbf {\bibinfo {volume} {21}},\ \bibinfo {pages} {043004}
  (\bibinfo {year} {2019})}\BibitemShut {NoStop}%
\bibitem [{\citenamefont {Borgas}\ \emph {et~al.}(2004)\citenamefont {Borgas},
  \citenamefont {Sawford}, \citenamefont {Xu}, \citenamefont {Donzis},\ and\
  \citenamefont {Yeung}}]{BSXDY}%
  \BibitemOpen
  \bibfield  {author} {\bibinfo {author} {\bibfnamefont {M.~S.}\ \bibnamefont
  {Borgas}}, \bibinfo {author} {\bibfnamefont {B.~L.}\ \bibnamefont {Sawford}},
  \bibinfo {author} {\bibfnamefont {S.}~\bibnamefont {Xu}}, \bibinfo {author}
  {\bibfnamefont {D.~A.}\ \bibnamefont {Donzis}}, \ and\ \bibinfo {author}
  {\bibfnamefont {P.~K.}\ \bibnamefont {Yeung}},\ }\bibfield  {title} {\enquote
  {\bibinfo {title} {{High Schmidt number scalars in turbulence: structure
  functions and Lagrangian theory}},}\ }\href@noop {} {\ \textbf {\bibinfo
  {volume} {16}},\ \bibinfo {pages} {3888--3899} (\bibinfo {year}
  {2004})}\BibitemShut {NoStop}%
\bibitem [{\citenamefont {Yasuda}\ \emph {et~al.}(2020)\citenamefont {Yasuda},
  \citenamefont {Gotoh}, \citenamefont {Watanabe},\ and\ \citenamefont
  {Saito}}]{yasuda20}%
  \BibitemOpen
  \bibfield  {author} {\bibinfo {author} {\bibfnamefont {T.}~\bibnamefont
  {Yasuda}}, \bibinfo {author} {\bibfnamefont {T.}~\bibnamefont {Gotoh}},
  \bibinfo {author} {\bibfnamefont {T.}~\bibnamefont {Watanabe}}, \ and\
  \bibinfo {author} {\bibfnamefont {I.}~\bibnamefont {Saito}},\ }\bibfield
  {title} {\enquote {\bibinfo {title} {P\'eclet-number dependence of
  small-scale anisotropy of passive scalar fluctuations under a uniform mean
  gradient in isotropic turbulence},}\ }\href@noop {} {\bibfield  {journal}
  {\bibinfo  {journal} {J.~Fluid Mech.}\ }\textbf {\bibinfo {volume} {898}},\
  \bibinfo {pages} {A4} (\bibinfo {year} {2020})}\BibitemShut {NoStop}%
\bibitem [{\citenamefont {Shraiman}\ and\ \citenamefont {Siggia}(1994)}]{ss94}%
  \BibitemOpen
  \bibfield  {author} {\bibinfo {author} {\bibfnamefont {B.~I.}\ \bibnamefont
  {Shraiman}}\ and\ \bibinfo {author} {\bibfnamefont {E.~D.}\ \bibnamefont
  {Siggia}},\ }\bibfield  {title} {\enquote {\bibinfo {title} {Lagrangian path
  integrals and fluctuations in random flow},}\ }\href@noop {} {\bibfield
  {journal} {\bibinfo  {journal} {Phys.~Rev.~E}\ }\textbf {\bibinfo {volume}
  {49}},\ \bibinfo {pages} {2912} (\bibinfo {year} {1994})}\BibitemShut
  {NoStop}%
\end{thebibliography}

%

\end{document}